\documentclass[prl,twocolumn,twoside,preprintnumbers,superscriptaddress, nofootinbib]{revtex4-2}
\usepackage{physics}
\usepackage{amsmath}
\usepackage{amssymb}

\usepackage{hyperref}
\usepackage{enumitem}

\usepackage{xspace}
\usepackage{slashed}
\usepackage{braket}
\usepackage{leftindex}
\hypersetup{
    colorlinks=true,       % false: boxed links; true: colored links
    linkcolor=blue,          % color of internal links
    citecolor=blue,        % color of links to bibliography
    filecolor=blue,      % color of file links
    urlcolor=blue           % color of external links
}

\usepackage{graphicx}  %
\usepackage{bm}  %                                          
\usepackage{amsmath}   %
\usepackage{graphics}%, slashed, braket}
\usepackage{color}
\usepackage{colortbl}
\usepackage{ulem}
\usepackage{hyperref}
\usepackage{xcolor}
\usepackage{comment}
\usepackage{enumitem}

\newcommand{\Ocal}[0]{\mathcal{O}}

\renewcommand{\vec}[1]{{\mathbf #1}}

\newcommand{\dslash}[1]{#1 \llap{/\kern-0.5pt}}
\newcommand{\Dslash}[1]{#1 \llap{/\kern+1.5pt}}
\newcommand{\DDslash}[1]{#1 \llap{/\kern+2.3pt}}
\newcommand{\dslashh}[1]{#1 \llap{/\kern+1pt}}

\newcommand{\bea}{\begin{eqnarray}}
\newcommand{\eea}{\end{eqnarray}}
\newcommand{\be}{\begin{equation}}
\newcommand{\ee}{\end{equation}}
\newcommand{\bma}{\begin{pmatrix}}
\newcommand{\ema}{\end{pmatrix}}

\newcommand{\agl}[2]{\langle#1\, #2 \rangle}

\newcommand{\txt}[1]{\textup{#1}}
\allowdisplaybreaks[1]

\begin{document}
%\preprint{}

\title{Anomalous Dimensions 
%for Effective Field Theories 
via on-shell Methods:\\ 
%General 
Operator Mixing and Leading Mass Effects
%Multiple Operator Insertions and Leading Mass Effects
}

\author{L.~C. Bresciani}
\affiliation{Dipartimento di Fisica e Astronomia `G. Galilei', Universit\`a di Padova, Italy}
\affiliation{Istituto Nazionale Fisica Nucleare, Sezione di Padova, I--35131 Padova, Italy}
\author{G. Levati}
\affiliation{Dipartimento di Fisica e Astronomia `G. Galilei', Universit\`a di Padova, Italy}
\affiliation{Istituto Nazionale Fisica Nucleare, Sezione di Padova, I--35131 Padova, Italy}
\author{P. Mastrolia}
\affiliation{Dipartimento di Fisica e Astronomia `G. Galilei', Universit\`a di Padova, Italy}
\affiliation{Istituto Nazionale Fisica Nucleare, Sezione di Padova, I--35131 Padova, Italy}
\author{P. Paradisi}
\affiliation{Dipartimento di Fisica e Astronomia `G. Galilei', Universit\`a di Padova, Italy}
\affiliation{Istituto Nazionale Fisica Nucleare, Sezione di Padova, I--35131 Padova, Italy}
\begin{abstract}
We elaborate on the application of on-shell and unitarity-based methods for evaluating renormalization group coefficients, and generalize this framework to account for the mixing of operators with different dimensions and leading mass effects. 
We derive a master formula for anomalous dimensions stemming from the general structure of operator mixings, up to two-loop order, and show how the Higgs low-energy theorem can be exploited to include leading mass effects.
A few applications on the renormalization properties of popular effective field theories showcase the strength of 
the proposed approach, which drastically reduces the complexity of standard loop calculations.
Our results provide a powerful tool to interpret experimental measurements of low-energy observables, such as flavor violating processes or electric and magnetic dipole moments, as induced by new physics emerging above the electroweak scale.
\end{abstract}

\maketitle

%\section{I. Introduction} 
%
{\bf Introduction} The Standard Model (SM) of particle physics has passed unclashed several experimental tests in all its sectors. The lack for heavy 
new physics (NP) at the LHC has firmly established the SM as a very successful theory describing the fundamental interactions of Nature up to the TeV scale. However, it is a common belief that the SM has to be regarded as the low-energy description of a more fundamental theory emerging at a large, yet unknown, energy scale $\Lambda$. 
New interactions can be then described by an Effective Field Theory (EFT) containing non-renormalizable operators that are invariant under the SM gauge group. EFTs provide a very powerful and model-independent approach to NP which does not rely on the details of the underlying (unknown) high-energy theory but just on its symmetries.

Predictions for physical processes are obtained by evaluating matrix elements of the EFT Lagrangian at energy scales accessible by collider experiments. Therefore, the high-scale Lagrangian needs to be evolved from the scale $\Lambda$ down to the experimental scale 
$E \ll \Lambda$. Such a program can be carried out by computing the anomalous dimension matrix of the higher-dimension operators which control both the multiplicative renormalization of operators as well as their mixing effects. 
In particular, the latter provide important information on how experimental bounds from one operator impact the Wilson coefficients of other operators. This makes the evaluation of EFT anomalous dimension matrices a crucial ingredient for interpreting experimental results. 
%The exploration of NP effects at the scale $\Lambda$ can be systematically performed in a model-independent fashion by relying on Effective Field Theory (EFT) approaches which require the knowledge of just the symmetries but not the details of the underlying high-energy theory. 
%Scattering-amplitude methods have been successfully applied to a number of EFT studies including positivity constraints [48–50], non-interferences for NP amplitudes through helicity selection rules [51], non-renormalisation theorems between higher-dimensional operators [52–56], construction of operator bases [66–72] and massive amplitudes [73–81]. Moreover, they provide an alternative way of computing anomalous dimensions [57–65].
A systematic and comprehensive computation of the one-loop anomalous dimension matrix has been carried out for a number of relevant EFTs,
such as the Standard Model EFT (SMEFT)~\cite{Jenkins:2013zja,Jenkins:2013wua,Alonso:2013hga,Jenkins:2017dyc}  or the axion-like particle EFT~\cite{Bauer:2020jbp,Chala:2020wvs}, exploiting diagrammatic and functional methods. 

% in its unbroken~\cite{Jenkins:2013zja,Jenkins:2013wua,Alonso:2013hga} as well as broken\cite{Jenkins:2017dyc} phase or axion-like-particle EFTs~\cite{Bauer:2020jbp,Chala:2020wvs}, exploiting diagrammatic and functional methods. 

Recently, the calculation of anomalous dimensions has been addressed also employing
%by making use of 
on-shell and unitarity-based techniques for scattering amplitudes~\cite{Caron-Huot:2016cwu,EliasMiro:2020tdv,Baratella:2020lzz,Jiang:2020mhe,Bern:2020ikv,Baratella:2020dvw,AccettulliHuber:2021uoa,EliasMiro:2021jgu,Baratella:2022nog,Machado:2022ozb}. 
One of the most intriguing feature of this approach is to make manifest 
hidden structures with the appearance of nontrivial zeros in the anomalous dimension matrix. The origin of these vanishing elements has been traced back to selection rules~\cite{Elias-Miro:2014eia}, helicity~\cite{Cheung:2015aba}, operator lengths~\cite{Bern:2019wie}, 
and angular momentum conservation~\cite{Jiang:2020rwz}.

Anomalous dimensions can be extracted from the ultraviolet divergent 
part of amplitudes exploiting the generalized unitarity method~\cite{Bern:1994zx,Bern:1994cg,Britto:2004nc,Britto:2005ha,Britto:2006sj,Mastrolia:2009dr} for assembling scattering amplitudes from their unitarity cuts (see also~\cite{Ellis:2011cr}, for review). 
Therefore, unitarity cuts give a direct access to the renormalization-scale dependence. 
%After subtracting infrared singularities, the renormalization-scale dependence can be read off from the arguments of logarithms [16]. 
In Ref.~\cite{Caron-Huot:2016cwu}, it was remarkably observed that anomalous dimensions can be directly related to unitarity cuts. 
In particular, the discontinuities of form factors of EFT 
operators can be calculated via phase-space integrals and are related 
to the corresponding anomalous dimensions. This method has been shown to 
be particularly effective for computing anomalous dimensions at 
two-loop order~\cite{Bern:2020ikv}.

So far, the method proposed in Ref.~\cite{Caron-Huot:2016cwu}, 
has been applied to derive the anomalous dimensions 
of non-renormalizable massless theories including only
mixing effects among operators of the same dimension. 
However, the non-trivial inclusion of mixings among operators 
with different dimensions and leading mass effects, which are 
of paramount importance in many popular EFT extensions of the SM, has not been discussed so far in this framework. The main motivation of this Letter is to fill this gap. In particular, we generalize the method of Ref.~\cite{Caron-Huot:2016cwu} providing a master formula 
which includes the most general operator mixing contributions up 
to two-loop order. Moreover, we show how to include leading 
mass effects, still working in the massless limit, by exploiting 
the Higgs low-energy theorem~\cite{Ellis:1975ap,Shifman:1979eb}. 
Few applications of our methods are illustrated by means of the renormalization of the axion-like particle EFT~\cite{Georgi:1986df,Marciano:2016yhf,DiLuzio:2020oah} 
and the low-energy EFT of the SM below 
the electroweak scale (LEFT)~\cite{Jenkins:2017dyc}.
\\
%The paper is organized as follows. In Section II, we summarize the method of~\cite{Caron-Huot:2016cwu}, while we provide its generalization to include general operator mixings and leading mass effects in Sections III and IV, respectively. In Section V, we explain our methods evaluating few anomalous dimensions of popular EFTs. Section VI is dedicated to our conclusions.

%\section{II. The method of form factors}
{\bf The method of form factors} In this Section, we review the method of Ref.~\cite{Caron-Huot:2016cwu}.
The fundamental objects we deal with are the form factors of local gauge-invariant operators $\mathcal{O}_i$ of the Lagrangian
%$\mathcal{L}_{\text{EFT}} = \sum_{n>4} \frac{c_n}{\Lambda^{n-4}}\, \mathcal{O}_i^{(n)}$.
$\mathcal L_{\text{EFT}} = \sum_i c_i \, \mathcal O_i / \Lambda^{[\mathcal O_i]-4}$.
These form factors are defined as
\begin{equation}
    F_i(\vec n;q)= \frac{1}{\Lambda^{[\mathcal O_i]-4}}\mel*{\vec n}{\mathcal O_i(q)}{0}\,,
\end{equation}
which is a matrix element between an outgoing on-shell state $\bra{\vec n}=\bra{1^{h_1},\dotsc,n^{h_n}}$ and an operator $\mathcal O_i$ that 
injects an additional off-shell momentum $q$.
In dimensional regularization, form factors depend
%acquire a dependence 
on the renormalization scale $\mu$ and satisfy 
the Callan-Symanzik equation
\begin{equation}
\label{eq:CSEq}
    \left(\delta_{ij} \mu\frac{\partial}{\partial \mu}+ \frac{\partial \beta_i}{\partial c_j}-\delta_{ij}\gamma_{i,\txt{IR}} +\delta_{ij}\beta_g\frac{\partial}{\partial g}\right)F_i=   0\,,
\end{equation}
where $g$ collectively denotes the couplings related to the renormalizable operators of our Lagrangian, while $\gamma_{i,\txt{IR}}$ is the infrared anomalous dimension. 
The renormalization of the operator $\mathcal{O}_i$ induced by 
$\mathcal{O}_j$ is described by
\begin{equation}
\beta_i(\{c_k\}) \equiv \mu \frac{\text{d} c_i}{ \text{d} \mu} = \gamma_{i\leftarrow j} \, c_j \, 
%\GL{ + \, \mathcal{O}(c_ic_j)?}\,,
\end{equation}
where $c_i$ are the Wilson coefficients of the effective Lagrangian $\mathcal{L}_{\text{EFT}}$.

Exploiting the analyticity of form factors, unitarity and the CPT theorem, it can be shown that an elegant relation exists linking the action of the dilatation operator ($D$)
%
%\begin{equation}
%    D = \sum_{\txt{all particles } i}p_i\cdot \frac{\partial}{\partial p_i}
%\end{equation}
%
to the action of the $S$-matrix ($S$) on form factors~\cite{Caron-Huot:2016cwu}:
\begin{equation}
\label{eq:FundamentalRelation}
    e^{-i \pi D}F_i^* = SF_i^*
\end{equation}
where $S=\mathbf{1} + i\mathcal{M}$ while 
$D = \sum_{i}p_i\cdot \partial/\partial p_i$
(the sum is over all particles $i$).
For a massless theory, in dimensional regularization, 
the latter reduces to $D \simeq -\mu \partial_\mu$. 

This allows to connect Eqs.~\eqref{eq:CSEq} and \eqref{eq:FundamentalRelation}, thus relating the 
infrared and ultraviolet anomalous dimensions to the 
scattering matrix phase when applied to a form factor. 
In particular, at one-loop order, it has been found that
\begin{equation}
\label{eq:masterformula}
\!\!\!\!
\left(
\frac{\partial \beta_i^{(1)}}{\partial c_j}
%\gamma_{i\leftarrow j}^{(1)}
-\delta_{ij}\gamma_{i,\txt{IR}}^{(1)} 
+\delta_{ij}\beta_g^{(1)}\frac{\partial}{\partial g}
\right)
F_i^{(0)} = -\frac{1}{\pi}(\mathcal M F_j)^{(1)}
\end{equation}
%
%which allows us to extract $\gamma_{i\leftarrow j}^{(1)}$ if we evaluate this equation at the Gaussian fixed point \footnote{IR anomalous dimensions can be either imported from standard computations or by making use of the method of form factors itself, provided that it is applied to any UV-safe local gauge-invariant operator. The choice in the latter case usually falls on the energy-momentum tensor for obvious reasons.}:
%
%\begin{equation}
%\!\!\!\!\!\left(\gamma_{i\leftarrow j}^{(1)}-\delta_{ij}\gamma_{i,\txt{IR}}^{(1)} +\delta_{ij}\beta_g^{(1)}\frac{\partial}{\partial g}\right)F_i|_{*}^{(0)} = -\frac{1}{\pi}(\mathcal M F_j)|_{*}^{(1)}
%\label{eq:master_formula}
%\end{equation}
%
where the right-hand side of Eq.~\eqref{eq:masterformula} corresponds to a sum over all one-loop two-particle unitarity cuts
%
\begin{comment}
\begin{equation}
\label{eq:MF1Loop}
\begin{split}
    (\mathcal M F_j)^{(1)} &= \sum_{k=2}^n \mathcal M^{(0)}_{2\to k}\otimes F_{n-k+2,j}^{(0)}\,. 
\end{split}
\end{equation}
\end{comment}
%
\begin{align}
\label{eq:MF1Loop}
    &\!\!(\mathcal M F_j)^{(1)}(1,\dots,n) = 
   \sum_{k=2}^n \sum_{\{\ell_1,\ell_2\}}\int d\text{LIPS}_2 
   \nonumber\\ 
   &\!\!\sum_{h_1,h_2}\!\! F_j^{(0)}\!(\ell_1^{h_1},\ell_2^{h_2},k \!+\! 1,\dots,n) \mathcal M^{(0)}\!(1,\dots,k;\ell_1^{h_1},\ell_2^{h_2}\!) ,
\end{align}
$\mathcal M(\vec n; \vec m)=\mel*{\vec n}{\mathcal M}{\vec m}$ and $d\text{LIPS}_2$ is the (two-particle) Lorentz invariant phase-space measure. The corresponding  cut-integral can be evaluated employing different parametrizations,  
by angular integration \cite{Caron-Huot:2016cwu,EliasMiro:2020tdv,Baratella:2020lzz,Bern:2020ikv,Baratella:2020dvw}, 
or via Stokes' theorem~\cite{Mastrolia:2009dr,Jiang:2020mhe}.
\\

{\bf General operator mixing}
%Multiple operator insertions 
The mixing among operators of different dimensions is required in many EFTs in order to capture the leading effects to several observables. 
This feature can be elegantly included within the method of form factors as we are going to discuss.

The crucial observation is that, in the neighborhood of the Gaussian 
fixed point ($*$), where $c_i=0$ $\forall i$, the renormalization group equations for the Wilson coefficients $c_i$ can be Taylor expanded as
\begin{equation}
\begin{split}
    \mu\dv[]{c_i}{\mu} &= \sum_{n> 0}
    %\sum_{j_1,\dotsc,j_n}
    \frac{1}{n!}\gamma_{i\leftarrow j_1,\dotsc,j_n}c_{j_1}\dotsm c_{j_n}\\
    &=
    %\sum_j
    \gamma_{i\leftarrow j}c_j +
    %\sum_{j,k}
    \frac{1}{2}\gamma_{i\leftarrow j,k}c_jc_k + \dotsb\,,
\end{split}
\end{equation}
where 
\begin{equation}
    \gamma_{i\leftarrow j_1, \dotsc, j_n} = \left.\frac{\partial^n \beta_i}{\partial c_{j_1}\dotsm \partial c_{j_n}}\right |_{*}
\end{equation}
has a perturbative expansion in the couplings of the leading order Lagrangian: $\gamma_{i\leftarrow j_1, \dotsc, j_n}= \sum_{\ell > 0}\gamma_{i\leftarrow j_1, \dotsc, j_n}^{(\ell)}$.

Focusing on the most relevant case of a double operator insertion, 
the key object we have to evaluate is $\gamma_{i\leftarrow j,k}$. 
A particularly convenient way to write it is the following
\begin{align}
    \gamma_{i\leftarrow j,k}&=\left. \frac{\partial^2 \beta_i}{\partial c_j \partial c_k}\right|_{*} = \left.\frac{\partial}{\partial c_k}\right|_{c_k=0} \!\!\left.\frac{\partial \beta_i}{\partial c_j}\right|_{*,c_k\neq 0} \,.
    \label{eq:gamma_ijk}
\end{align}
In fact, the last equality of Eq.~\eqref{eq:gamma_ijk}
enables us to generalize the master formula of Eq.~\eqref{eq:masterformula} by simply differentiating it with respect to a Wilson coefficient and then evaluating the result at the Gaussian fixed point, where $c_k = 0$.

At one-loop order, we obtain the following expression 
\begin{multline}
\label{eq:mastergammageneral}
    \left( \gamma_{i\leftarrow j,k}^{(1)} - \delta_{ij}\left.\frac{\partial \gamma^{(1)}_{i,\text{IR}}}{\partial c_k} \right|_{*}+\delta_{ij}\left.\frac{\partial \beta_g^{(1)}}{\partial c_k} \right|_{*}\frac{\partial}{\partial g} \right) F_i|_{*}^{(0)} = \\ -\frac{1}{\pi}\left.\frac{\partial}{\partial c_k}\right|_{c_k=0}(\mathcal M F_j)|_{*,c_k\neq 0}^{(1)} \,,
\end{multline}
which represents an important result of this Letter.
Since we are interested in mixing of operators with different dimensions, hereafter, we focus on the case where $j,k\neq i$ %Therefore, Eq.~\eqref{eq:mastergammageneral} reduces to
%However, from now on, we specialize to cases where $j,k\neq i$, which are by far the most relevant ones when EFTs are considered.
%Therefore, Eq.~\eqref{eq:mastergammageneral} reduces to
%
\begin{align}
\label{eq:mastergammaijk}
\gamma_{i\leftarrow j,k}^{(1)}F_i|_{*}^{(0)} &= -\frac{1}{\pi}\left.\frac{\partial}{\partial c_k}\right|_{c_k=0}(\mathcal M F_j)|_{*,c_k\neq 0}^{(1)} \,,
%\nonumber \\
%&= -\frac{1}{\pi}\left.\frac{\partial}{\partial c_k}\right|_{c_k=0}(\mathcal M F_j)|_{*,c_k\neq 0}^{(1)}\,.
\end{align}
which can be generalized at two-loop order by properly expanding Eq.~\eqref{eq:FundamentalRelation} at the desired order. We find
%
%The expression corresponding to Eq.~\eqref{eq:mastergammaijk} at two-loop order can be obtained as well, by properly expanding Eq.~\eqref{eq:FundamentalRelation} at the desired order. We find that
%
%
\begin{align}
\label{eq:mastergammaijk_2loop}
&\gamma_{i\leftarrow j,k}^{(2)} F_i|_{*}^{(0)} = -\frac{1}{\pi}\left.\frac{\partial}{\partial c_k}\right|_{c_k=0}%\!\!\!\!\!\!\!
(\text{Re}\mathcal M \, \text{Re} F_j)|_{*,c_k\neq 0}^{(2)} \nonumber \\
%&-  \left[\Delta \gamma_{ij}^{(1)}\! + \delta_{ij} \beta^{(1)} \frac{\partial}{\partial g}\right]\!\left.\frac{\partial}{\partial c_k}\right|_{c_k=0} \!\!\!\!\!\!\!\!\!\!\!\!\text{Re}\,F_j|_{*,c_k\neq 0}^{(1)} \nonumber\\
& - \gamma_{i\leftarrow j}^{(1)}\left.\frac{\partial}{\partial c_k}\right|_{c_k=0} \text{Re}F_i|_{*,c_k\neq 0}^{(1)} 
- \gamma_{i\leftarrow j,k}^{(1)} \text{Re}F_i|_{*}^{(1)}\,.
\end{align}
%
%
%\begin{align}
%\label{eq:mastergammaijk_2loop}
%\gamma_{i\leftarrow j,k}^{(2)} F_i|_{*}^{(0)} = &-\frac{1}{\pi}\left.\frac{\partial}{\partial c_k}\right|_{c_k=0}%\!\!\!\!\!\!\!
%(\text{Re}\mathcal M \, \text{Re} F_j)|_{*,c_k\neq 0}^{(2)} \nonumber \\
%&-  \left[\Delta \gamma_{ij}^{(1)}\! + \delta_{ij} \beta^{(1)} \frac{\partial}{\partial g}\right]\!\left.\frac{\partial}{\partial c_k}\right|_{c_k=0} \!\!\!\!\!\!\!\!\!\!\!\!\text{Re}\,F_j|_{*,c_k\neq 0}^{(1)} \nonumber\\
%& - \gamma_{i\leftarrow j}^{(1)}\left.\frac{\partial}{\partial c_k}\right|_{c_k=0} \text{Re}F_i|_{*,c_k\neq 0}^{(1)} \nonumber\\
%& - \gamma_{i\leftarrow j,k}^{(1)} \text{Re}F_i|_{*}^{(1)}\,.
%\end{align}
%
The extension of Eqs.~\eqref{eq:mastergammageneral}, \eqref{eq:mastergammaijk} and \eqref{eq:mastergammaijk_2loop} 
with multiple operator insertions 
$\gamma_{i\leftarrow j_1, \dotsc, j_n}$ is straightforward.
\\

{\bf Leading mass effects}
As a natural consequence of operator mixing, chirality-violating and preserving operators do generally mix under the renormalization flow.
In case of EFTs defined below the electroweak scale, the required chirality flip proceeds through a fermion mass insertion. However, such a mass dependence cannot be directly implemented in the massless method of Ref.~\cite{Caron-Huot:2016cwu}. 

In order to circumvent this issue, we rely on the low-energy Higgs theorem~\cite{Ellis:1975ap,Shifman:1979eb}, which was originally introduced to estimate the properties of a light Higgs boson in analogy to how soft-pion theorems are used to study low-energy pion interactions~\cite{Adler:1964um,Cheung:2021yog,Balkin:2021dko,Bertuzzo:2023slg}. 

\begin{figure}[h]
\vspace{10pt}
\centering
\includegraphics[width=.9\linewidth]{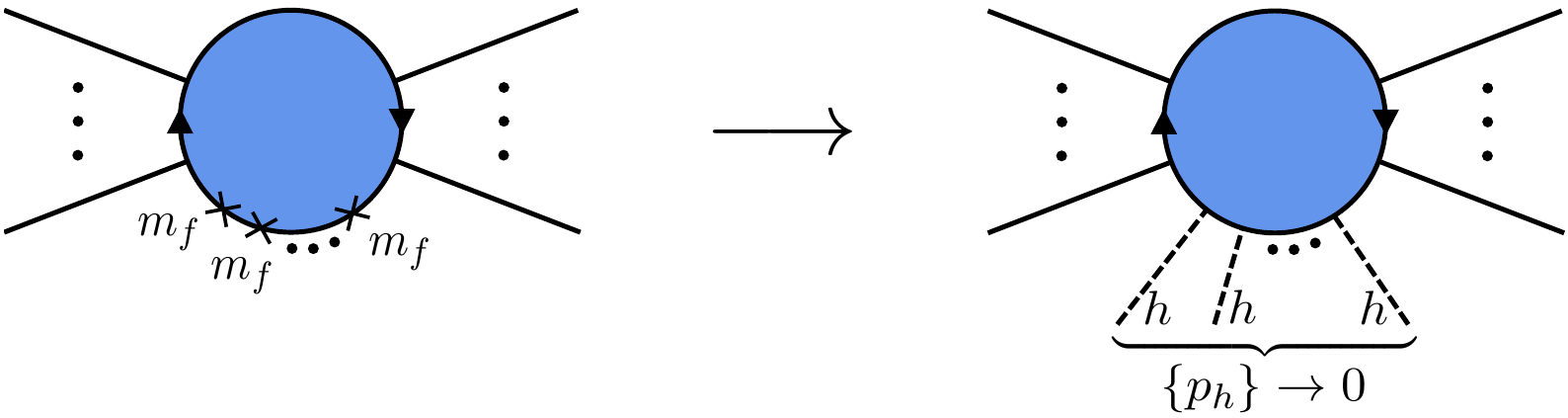}
\vspace{-10pt}
\caption{Diagrammatic representation of our method for massifying amplitudes based on the low-energy Higgs theorem. 
%Arrows signal the flow of chirality.
}
\label{fig:HMs}
\end{figure}

The key observation is that the fermionic Higgs interactions in the SM can be written in the following form:
\begin{align}
\mathcal{L}^{\textit{int}}_{H} = - \left( 1 + \frac{h}{v}\right) 
\sum_f m_f \bar f f 
\end{align}
where the sum is over all fermions in the theory. 
In the limit where the Higgs field $h$ has a vanishing 
four-momentum, $p_h\to 0$, $h$ becomes a constant field 
and its effect is equivalent to redefining all mass 
parameters as $m_f \to m_f \,(1 + h/v)$.
This immediately implies the following low-energy theorem~\cite{Ellis:1975ap,Shifman:1979eb}
\begin{align}
\!\!\!\lim_{\{p_{h}\} \to 0}\!\!\mathcal{M}(A \!\to\! B 
\!+\! Nh) \!=\! 
\sum_f \!\frac{m^N_f}{v^N} \frac{\partial^N}{\partial m_f^N} \mathcal{M}(A \!\to\! B)
\end{align}
relating the amplitudes of two processes differing by $N$ insertions of zero momentum Higgs bosons. 
%
%The key idea consists in promoting fermionic masses to dynamical massless spurionic fields, whose vacuum expectation value is eventually set to the mass of the fermions. This is reminiscent of what happens in the Higgs mechanism, where fermion masses and Yukawa interactions are one-to-one related via the term $m_f (1 + h/v) \bar{f}f$. 
%
In practice, whenever an amplitude requires $N$ fermion mass insertions not to vanish, we consider an equivalent amplitude entailing $N$ extra massless Higgs fields, see Fig.~\ref{fig:HMs}.

%%%%%%%%%%%%%%%%%%%%%%%%%%%%%%%%%%%%%%%%%%

%%%%%%%%%%%%%%%%%%%%%%%%%%%%%%%%%%%%%%%%%%

The number $N$ can be determined as follows. By dimensional 
analysis we argue that whenever the anomalous dimension 
$\gamma_{i\leftarrow j_1, \dotsc, j_n}$ vanishes in the 
limit of massless fermions, possible mass effects must 
be of order $(m_f/\Lambda)^N$, where
\begin{equation}
\label{eq:Sup_Div_Deg}
N = 4-[\mathcal{O}_{i}] +\sum^n_{k=1} ([\mathcal{O}_{j_k}]-4)%\geq 0\,
\end{equation}
is the superficial degree of divergence associated with the loop diagram under consideration (see the Appendix for a derivation).
%~\footnote{See Supplemental Material for a derivation of Eq.~\eqref{eq:Sup_Div_Deg}}. 
Notice that the number of needed Higgs insertions coincides with the superficial degree of divergence in Eq.~\eqref{eq:Sup_Div_Deg} because scaleless integrals vanish in dimensional regularization.

For $N < 0$, $\gamma_{i\leftarrow j_1, \dotsc, j_n}$ is trivially zero.
For $N \geq 0$, the anomalous dimension
%$\gamma_{i\leftarrow j_1, \dotsc, j_n}$ 
is obtained by renormalizing the operator $\mathcal{O}^{Nh}_i= (h/v)^N\mathcal{O}_i/N!$
instead of $\mathcal{O}_i$.

\begin{figure}[h]
%\vspace{10pt}
\centering
\includegraphics[width=\linewidth]{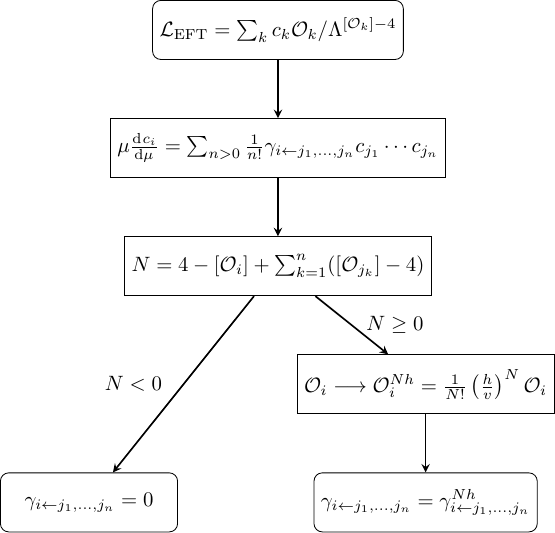}
%\vspace{-10pt}
\caption{Algorithm for the computation of anomalous dimensions that  require $N$ fermionic mass insertions not to vanish. 
}
\label{fig:flowchart}
\end{figure}

The procedure outlined before is summarized in the algorithm in Fig.~\ref{fig:flowchart}.

%%%%%%%%%%%%%%%%%%%%%%%%%%%%%%%%%%%%%%%%%%
%%%%%%%%%%%%%%%%%%%%%%%%%%%%%%%%%%%%%%%%%%

Remarkably, the approximation of setting the Higgs mass to 
zero is justified in our study since we are interested in 
the evaluation of anomalous dimensions that are related to 
the ultraviolet properties of a theory.
\\

%%%%%%%%%%%%%%%%%%%%%%%%%%%%%%%%%%%%%%%%%%
\begin{figure*}[ht]
%\begin{figure*}[ht]
%\vspace{-60 pt}
\includegraphics[width=1.0\linewidth]{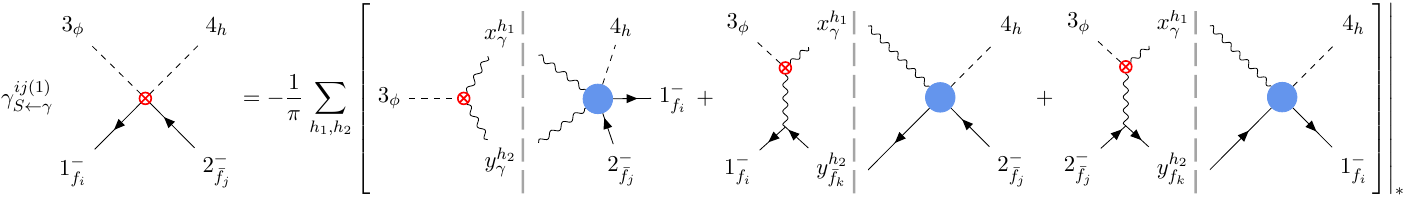}
%\vspace{-60pt}
\caption{Diagrams contributing to the renormalization of the $\mathcal O_{S}^h=(h/v)\phi \bar{f}f $ operator as induced by the $\mathcal O_\gamma=\phi FF$ one.
Here and in the following figures, red and light-blue blobs refer to form factors and amplitudes, respectively.}
\label{fig:CFDs}
\end{figure*}
%%%%%%%%%%%%%%%%%%%%%%%%%%%%%%%%%%%%%%%%%%

{\bf Phenomenological applications} 
In this Section, we discuss some applications of our results. First, we illustrate how to separately deal with either the mixing of operators with different dimensions or leading mass effects in axion-like particle EFT~\cite{Bauer:2020jbp,Chala:2020wvs,DiLuzio:2020oah}. 
Then, we show how to treat these effects simultaneously 
in the context of the LEFT~\cite{Jenkins:2017dyc}.
The most general effective Lagrangian describing 
axion-like particle (ALP) interactions with SM fields reads~\cite{Georgi:1986df,Marciano:2016yhf,DiLuzio:2020oah}
\begin{align}
    \mathcal{L}_\phi &= 
    \frac{\tilde{\mathcal{C}}_\gamma}{\Lambda}\,\phi\, F\tilde{F}
    + \frac{\tilde{\mathcal{C}}_g}{\Lambda}\,\phi\, G\tilde{G} 
    + i\, \mathcal{Y}^{ij}_P\,\phi\, \bar{f}_i \gamma_5 f_j 
    \nonumber\\
    & + \frac{\mathcal{C}_\gamma}{\Lambda}\,\phi\, FF
    + \frac{\mathcal{C}_g}{\Lambda}\,\phi\, GG 
    + \mathcal{Y}^{ij}_S\,\phi\, \bar{f}_i f_j \, ,
\end{align}
where $\phi$ is the ALP field, $\Lambda \gg v \approx 246$ GeV is the EFT cutoff scale, and $f\in \{e,u,d\}$ denotes SM fermions in the mass basis. Moreover, $G$ and $F$ are the QCD and QED field-strength tensors, while $\tilde{G}$ and $\tilde{F}$ are their duals. 

The need for an appropriate treatment of leading mass effects is evident in the renormalization of the operator $\mathcal O_S = \phi \bar{f}f$ as induced by $\mathcal O_\gamma= \phi FF$. Indeed, due to the chirality mismatch between the two operators, a mass insertion would be necessary to obtain a non-null result. This situation can be handled
%, as discussed before, 
by renormalizing the operator $\mathcal O_{S}^h=(h/v)\phi \bar f f$ instead of $\phi \bar f f$,
%--- divided by $v$ in order to preserve the right mass dimension --- and write
%
\begin{equation}
% \frac{1}{v}
  \gamma_{S\leftarrow \gamma}^{ij(1)}
%  \gamma_{hS}^{ij(1)}
  F_{S}^h|_{*}^{(0)}=-\frac{1}{\pi}(\mathcal M F_{\gamma})|_{*}^{(1)}\,,
\end{equation}
which is represented in Fig.~\ref{fig:CFDs}.
Summing up the three contributions 
of Fig.~\ref{fig:CFDs}, we obtain~\footnote{
We adopt the spinor-helicity formalism, where the spinor inner product is defined as $\agl{i}{j}=\epsilon_{\alpha\beta}\lambda_i^\alpha \lambda_j^\beta$, where $\epsilon_{\alpha\beta}$ is the $SL(2,\mathbb C)$ invariant Levi-Civita tensor, $\bar \sigma_\mu^{\dot\alpha\alpha} p_i^\mu= \tilde\lambda_i^{\dot\alpha}\lambda_i^\alpha$ and $\bar\sigma_\mu^{\dot\alpha\alpha}=
(\mathbf{1},\bm{\sigma})^{\dot \alpha\alpha}$.}:
\begin{equation}
\!\!\!\! (\mathcal M F_\gamma)|_{*}^{(1)}(1^-_{f_i},2^-_{\bar f_j},3_\phi,4_h) = -\frac{3}{2\pi\Lambda}
    \frac{m_i}{v}\delta^{ij}e^2 Q_f^2 \agl{1}{2}.
\end{equation}
Using 
\begin{equation}
F_{S}^h|_{*}^{(0)}(1^-_{f_i},2^-_{\bar f_j},3_\phi,4_h) = \frac{1}{v}\agl{1}{2}
\end{equation}
one can then find 
the sought-after result 
\begin{equation}
\gamma_{S\leftarrow \gamma}^{ij(1)} 
=\frac{3}{2\pi^2}
%m_i 
\frac{m_i}{\Lambda}\delta^{ij}e^2 Q_f^2 \,, 
\end{equation}
which agrees with Refs.~\cite{Bauer:2020jbp,Chala:2020wvs,DiLuzio:2020oah}. 

%\subsubsection{Operator mixing}

%%%%%%%%%%%%%%%%%%%%%%%%%%%%%%%%%%%%%%%%%%
\begin{figure}[h]
%\vspace{-30pt}
\includegraphics[width=1\linewidth]{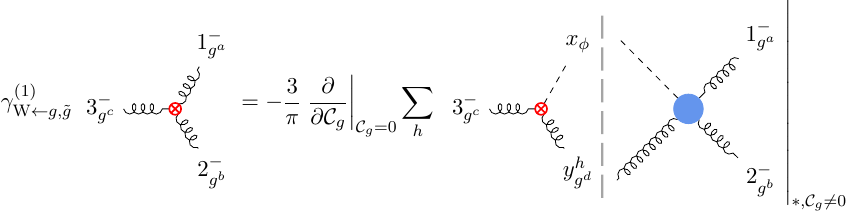}
\vspace{-20pt}
\caption{Diagrams contributing to the renormalization of the Weinberg operator $\mathcal O_{\rm W} = f^{abc}G^a_{\mu\rho}G^{b\;\;\rho}_{\;\nu}\tilde G^{c\,\mu\nu}/3$ as induced by the 
operators $\mathcal O_g=\phi\, GG$ and $\mathcal O_{\tilde g}=\phi\, G \tilde{G}$. The factor of $3$ 
on the right-hand side stems from the permutation of the gluons.
}
\label{fig:WOPs}
\end{figure}
%%%%%%%%%%%%%%%%%%%%%%%%%%%%%%%%%%%%%%%%%%

%%%%%%%%%%%%%%%%%%%%%%%%%%%%%%%%%%%%%%%%%%
%\begin{figure*}
\begin{figure*}[ht!]
%\vspace{-60 pt}
\includegraphics[width=1.0\linewidth]{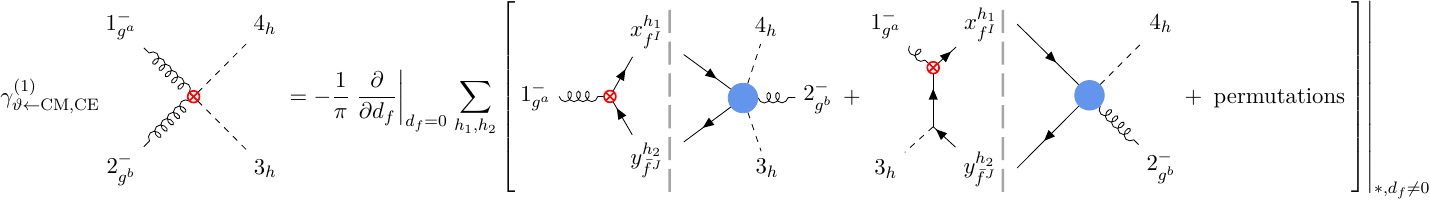}
%\vspace{-60pt}
\caption{Diagrams contributing to the renormalization of the operator $\mathcal O_{\vartheta}^{2h} = (h^2/2v^2) G\tilde G$ as induced by $\mathcal O_{\text{CM}}$ and $\mathcal O_{\text{CE}}$.}
\label{fig:gamma_theta}
\end{figure*}
%%%%%%%%%%%%%%%%%%%%%%%%%%%%%%%%%%%%%%%%%%

As an example of the impact that two insertions of lower-dimensional operators can have on a higher-dimensional one, 
we will consider here the generation of the Weinberg operator $\mathcal O_{\rm W} =f^{abc}G^a_{\mu\rho}G^{b\;\;\rho}_{\;\nu}\tilde G^{c\,\mu\nu}/3$ as induced by the simultaneous presence of the ALP-gluon couplings $\mathcal O_g=\phi GG$ 
and $\mathcal O_{\tilde g}=\phi G\tilde{G}$. According to the results of the previous Section, the anomalous dimension matrix element for $\mathcal O_{\rm W}$ can be extracted by evaluating  
\begin{equation}\label{eq:gammagg}
    \gamma^{(1)}_{\txt{W}\leftarrow g,\tilde g}F_{\rm W}|_{*}^{(0)}=-\frac{1}{\pi}\left. \frac{\partial}{\partial \mathcal C_g} \right|_{\mathcal C_g = 0} (\mathcal M F_{\tilde g})|_{*,\mathcal C_g \neq  0}^{(1)}
\end{equation}
where 
\begin{equation}
F_{\rm W}|_{*}^{(0)}(1^-_{g^a},2^-_{g^b},3^-_{g^c}) = \frac{\sqrt 2}{\Lambda^2} f^{abc} \agl{1}{2}\agl{2}{3}\agl{3}{1}
\end{equation}
is the form factor corresponding to the Weinberg operator for 
three negative-helicity gluons. Remarkably, as shown in Fig.~\ref{fig:WOPs}, we need to calculate just one 
contribution, in contrast with the standard diagrammatic method 
which requires twelve one-loop diagrams. We find
\begin{equation}
    (\mathcal M F_{\tilde g})|_{*,\mathcal C_g \neq  0}^{(1)}(1^-_{g^a},2^-_{g^b},3^-_{g^c}) = \frac{3\sqrt{2}}{\pi\Lambda^2} g_s\mathcal  C_gf^{abc} \agl{1}{2}\agl{2}{3} \agl{3}{1}
\end{equation}
yielding in turn
\begin{equation}
    \gamma^{(1)}_{\txt{W}\leftarrow g,\tilde g} = -\frac{3g_s}{\pi^2}\,,
\end{equation}
in agreement with the literature~\cite{DiLuzio:2020oah}.

In our last example, we show how to treat simultaneously 
the methods of the previous sections, by considering the renormalization 
of the QCD theta term, which is associated with the dimension-4 pseudoscalar density $\mathcal O_\vartheta=G\tilde G$. 
Focusing on dipole operators, the relevant dimension-5 LEFT
Lagrangian reads
\begin{equation}
\begin{split}
\mathcal L^{(5)} \supset \frac{a_f}{\Lambda} \,\mathcal O_{\rm CM}+ 
\frac{d_f}{\Lambda} \,\mathcal O_{\rm CE}\,,
\end{split}
\end{equation}
where $\mathcal O_{\rm CM}$ is the chromomagnetic operator while
$\mathcal O_{\rm CE}$ is the chromoelectric one, defined as
\begin{equation}
    \mathcal O_{\rm CM} = \bar f \sigma^{\mu\nu} T^a f G^a_{\mu\nu}\,,
~~~~
    \mathcal O_{\rm CE} = i\bar f \sigma^{\mu\nu}\gamma_5 T^a f G^a_{\mu\nu} \,.
\end{equation}
In the limit of massless fermions, the anomalous dimension
$\gamma_{\vartheta\leftarrow \text{CM},\text{CE}}$ vanishes. 
According to the previous discussion on the low-energy 
Higgs theorem, possible mass effects must be of order 
$(m_f/\Lambda)^2$.
%, as one can infer by dimensional analysis considerations.
%The first non-vanishing contribution to $\gamma^{(1)}_{\vartheta\leftarrow \text{CM},\text{CE}}$ is of order $m^2_f/\Lambda^2$, as it is inferred by dimensional analysis considerations.
Therefore, the required double Higgs insertion can be accounted 
for by introducing the $\mathcal{O}^{2h}_{\vartheta}=(h^2/2v^2) G\tilde G$ operator, as shown in Fig.~\ref{fig:gamma_theta}. 
Then, $\gamma^{(1)}_{\vartheta\leftarrow \text{CM},\text{CE}}$ can be extracted from
%The anomalous dimension of interest can then be extracted from
%
\begin{equation}\label{eq:gamma_thetaCMCE}
%    \frac{1}{v^2}
\!\! \gamma^{(1)}_{\vartheta\leftarrow \text{CM},\text{CE}} F^{2h}_{\vartheta}|_{*}^{(0)} = -\frac{1}{\pi}\left.\frac{\partial}{\partial d_f}\right|_{d_f=0}\!(\mathcal M F_{\text{CM}})|_{*,d_f\neq 0}^{(1)}
\end{equation}
where
\begin{equation}
F_{\vartheta}^{2h}|_{*}^{(0)}(1^-_{g^a},2^-_{g^b},3_h,4_h)=-\frac{2i}{v^2}\delta^{ab}\agl{1}{2}^2\,.
\end{equation}
By taking into account permutations of external particles, namely $(1^-_{g^a} \leftrightarrow 2^-_{g^b})$ and $(3_h \leftrightarrow 4_h)$, we obtain
\begin{equation}
    (\mathcal M F_{\text{CM}})|_{*,d_f\neq 0}^{(1)}(1^-_{g^a},2^-_{g^b},3_h,4_h) = 
    \frac{i}{\pi\Lambda^2}\frac{m_f^2}{v^2}d_f\delta^{ab} \agl{1}{2}^2\,.
\end{equation}
Inserting the above result in Eq.~\eqref{eq:gamma_thetaCMCE}, 
we find
\begin{equation}
    \gamma^{(1)}_{\vartheta\leftarrow \text{CM},\text{CE}} = \frac{m_f^2}{2\pi^2\Lambda^2}\,,
\end{equation}
in agreement with Ref.~\cite{Jenkins:2017dyc}.
More extensive applications are deferred to a companion study \cite{Bresciani:longpaper}. \\

{\bf Conclusions}
We have generalized the application of on-shell and unitarity-based methods for evaluating renormalization group coefficients, to account for the mixing of operators with different dimensions and leading mass effects, which play a fundamental role in the renormalization program of several effective field theories. 

In particular, we have derived a master formula accounting for operator 
mixings up to two-loop order, and shown how to include leading mass effects,
relying on the Higgs low-energy theorem. Our findings have been validated by reproducing well-established results of the literature, relative to popular effective field theories.

Our results can be applied to a number of new physics scenarios, 
defined above the TeV scale, in order to analyze their impact on 
low-energy observables (like flavor violating processes, 
electric and magnetic dipole moments, etc.)~occurring at or 
below the GeV scale.
Such a large separation of scales demands for the inclusion of 
running effects at two-loop order to obtain sensible predictions. 
While this is a very challenging task when approached with standard techniques, 
on-shell and unitarity-based methods offer a simpler, more efficient 
and elegant way to reach this goal. 
Our work may constitute an additional milestone for progressing along this direction.

%\bigskip
\vspace{5pt}

{\it Acknowledgments.} 
We thank S. De Angelis for many interesting discussions
and G. Brunello and M.K. Mandal for valuable checks.
This work received funding from the European Union’s Horizon 2020 research and innovation programme under the Marie Sklodowska-Curie grant agreement n.~101086085 – ASYMMETRY, 
by the INFN Iniziativa Specifica APINE, and by the INFN Iniziativa Specifica AMPLITUDES. The work of PP is supported by the Italian Ministry of University and Research (MUR) via the PRIN 2022 project n. 2022K4B58X – AxionOrigins.

%\vspace{60pt}

\section{Supplemental material}

{\bf EFT and superficial degree of divergence}
An $n$-particle amplitude related to the operator $\Ocal_i$ 
and entailing $m$ interaction vertices with the operators $\Ocal_{j_1},\dotsc,\Ocal_{j_m}$, can be schematically written as
%An $n$-particle amplitude (out of which the operator $\Ocal_i$ is composed) entailing $m$ interaction vertices mediated by the operators $\Ocal_{j_1},\dotsc,\Ocal_{j_m}$, can be schematically written as
%
%\begin{small}
\begin{equation}
    \!\!\!\mathcal M_{n} \!= C_{j_1}\!\!\dotsm C_{j_m} \!\!
     \int \!\! d^4 k_1 \!\dotsm d^4 k_L \frac{N
     (\{k\},\!\{p\})}{D(\{k\},\!\{p\})} 
    \alpha_1\!\dotsm \alpha_{n} 
\end{equation}
%\end{small}
where $\{p\}$ are external momenta and:
\begin{itemize}
    \item $C_{j_1},\dotsc, C_{j_m}$ are the Wilson coefficients corresponding to the operators $\Ocal_{j_1},\dotsc,\Ocal_{j_m}$ 
    such that
%\begin{small}
    \begin{equation}
        [C_{j_k}]= 4 - [\Ocal_{j_k}]\,.
    \end{equation}        
%\end{small}

    \item The $L$-loop integral has mass dimension 
%\begin{small}
    \begin{equation}
    \bigg[\int d^4 k_1\dotsi d^4 k_L\, \frac{N(\{k\},\{p\})}{D(\{k\},\{p\})}\bigg]=
    D + n_\partial
    \end{equation}        
%\end{small}
    %
    where $D$ is the degree of divergence of 
    $\mathcal M_{n}$ and $n_\partial$ the number 
    of derivatives contained in $\Ocal_i$.

%%%%%%%%%%%%%%%%%%%%%%%%%%%%%%%%%%%%%%%%%%%%%%%%%%%%%%%%%%%%%%%%%%%%%

    \item Depending on the bosonic/fermionic nature of the particle, $\alpha_1,\dotsc, \alpha_{n} \in \{1,\epsilon_\mu,\epsilon_\mu^*,u,v,\bar u, \bar v,\dotsc\}$. In the fermionic case, writing the 
    Dirac field as
    %\begin{small}
    \begin{equation}
     \Psi(x)\! =\! \int\! \frac{d^3k}{(2\pi)^3}  \frac{1}{\sqrt{2E_k}} 
     \left(a_k u \,e^{-ik\cdot x} \!+ 
     b^\dagger_k v \,e^{ik\cdot x}\right)
    \end{equation}
    %\end{small}
%
    and taking the quantization condition $\{a_k,a^\dagger_{k'}\}=\{b_k,b^\dagger_{k'}\}=(2\pi)^3\delta^{(3)}(k-k')$, 
    one finds that $[a]=[b]=-3/2$ and 
    $[u]=[v]\equiv [\alpha_{\text{fer}}]=1/2$. Similarly, one can also find that $[\alpha_{\text{bos}}] = 0$ .
    %One can always choose their mass dimension to be $[\Psi_{\text{bos}}] = 0$ and $[\Psi_{\text{fer}}] = \frac{1}{2}$.
\end{itemize}
By taking into account the above points, we find that
%\begin{small}
\begin{equation}\label{eq:massM1}
    [\mathcal M_n] = \sum_{k=1}^m(4-[\Ocal_{j_k}]) + n_\partial + D +\frac{1}{2}n_{\text{fer}}\,.
\end{equation}    
%\end{small}
%
The mass dimension of $\mathcal M_n$ can be inferred also from the $S$-matrix element
%
%\begin{small}
    \begin{equation}
    \mel{0}{S}{n} = (2\pi)^4 \, \delta^{(4)}(p_1+\dotsb+p_n) \, i \mathcal M_n
\end{equation}
%\end{small}
%
where $[S]=0$ and $[\ket{n}] = [(\sqrt{2E} a^\dagger)^n \ket{0}] = -n$ with $n = n_{\text{fer}} + n_{\text{bos}}$, resulting in
%
%\begin{small}
\begin{equation}\label{eq:massM2}
    [\mathcal M_{n}] = 4 - n\,.
\end{equation}
%\end{small}
%
From Eqs.~\eqref{eq:massM1} and \eqref{eq:massM2} 
and taking into account that
%
%\begin{small}
    \begin{equation}
    [\mathcal O_i] = n_{\text{bos}} + \frac{3}{2}n_{\text{fer}} + n_\partial\,,
\end{equation}
%\end{small}
%
it finally follows that
%\begin{small}
    \begin{equation}
    D= 4 - [\mathcal O_i] + \sum_{k=1}^m([\Ocal_{j_k}]-4)\,. 
\end{equation}
%\end{small}

%


\vspace{5pt}

\begin{thebibliography}{99}

%\cite{Jenkins:2013zja}
\bibitem{Jenkins:2013zja}
E.~E.~Jenkins, A.~V.~Manohar and M.~Trott,
%``Renormalization Group Evolution of the Standard Model Dimension Six Operators I: Formalism and lambda Dependence,''
JHEP \textbf{10} (2013), 087.
%doi:10.1007/JHEP10(2013)087
%[arXiv:1308.2627 [hep-ph]]
%555 citations counted in INSPIRE as of 28 Nov 2023
%
%\cite{Jenkins:2013wua}
\bibitem{Jenkins:2013wua}
E.~E.~Jenkins, A.~V.~Manohar and M.~Trott,
%``Renormalization Group Evolution of the Standard Model Dimension Six Operators II: Yukawa Dependence,''
JHEP \textbf{01} (2014), 035.
%doi:10.1007/JHEP01(2014)035
%[arXiv:1310.4838 [hep-ph]]
%560 citations counted in INSPIRE as of 28 Nov 2023
%
%\cite{Alonso:2013hga}
\bibitem{Alonso:2013hga}
R.~Alonso, E.~E.~Jenkins, A.~V.~Manohar and M.~Trott,
``%Renormalization Group Evolution of the Standard Model Dimension Six Operators III: Gauge Coupling Dependence and Phenomenology,''
JHEP \textbf{04} (2014), 159.
%doi:10.1007/JHEP04(2014)159
%[arXiv:1312.2014 [hep-ph]]
%723 citations counted in INSPIRE as of 28 Nov 2023
%
%\cite{Jenkins:2017dyc}
\bibitem{Jenkins:2017dyc}
E.~E.~Jenkins, A.~V.~Manohar and P.~Stoffer,
%``Low-Energy Effective Field Theory below the Electroweak Scale: Anomalous Dimensions,''
JHEP \textbf{01} (2018), 084.
%doi:10.1007/JHEP01(2018)084
%[arXiv:1711.05270 [hep-ph]]
%180 citations counted in INSPIRE as of 28 Nov 2023
%
%\cite{Bauer:2020jbp}
\bibitem{Bauer:2020jbp}
M.~Bauer, M.~Neubert, S.~Renner, M.~Schnubel and A.~Thamm,
%``The Low-Energy Effective Theory of Axions and ALPs,''
JHEP \textbf{04} (2021), 063.
%doi:10.1007/JHEP04(2021)063
%[arXiv:2012.12272 [hep-ph]]
%115 citations counted in INSPIRE as of 28 Nov 2023
%
%\cite{Chala:2020wvs}
\bibitem{Chala:2020wvs}
M.~Chala, G.~Guedes, M.~Ramos and J.~Santiago,
%``Running in the ALPs,''
Eur. Phys. J. C \textbf{81} (2021) no.2, 181.
%doi:10.1140/epjc/s10052-021-08968-2
%[arXiv:2012.09017 [hep-ph]]
%79 citations counted in INSPIRE as of 28 Nov 2023
%
%\cite{Caron-Huot:2016cwu}
\bibitem{Caron-Huot:2016cwu}
S.~Caron-Huot and M.~Wilhelm,
%``Renormalization group coefficients and the S-matrix,''
JHEP \textbf{12} (2016), 010.
%doi:10.1007/JHEP12(2016)010
%[arXiv:1607.06448 [hep-th]]
%43 citations counted in INSPIRE as of 26 Nov 2023
%
%\cite{EliasMiro:2020tdv}
\bibitem{EliasMiro:2020tdv}
J.~Elias Mir\'o, J.~Ingoldby and M.~Riembau,
%``EFT anomalous dimensions from the S-matrix,''
JHEP \textbf{09} (2020), 163.
%doi:10.1007/JHEP09(2020)163
%[arXiv:2005.06983 [hep-ph]]
%37 citations counted in INSPIRE as of 28 Nov 2023
%
%\cite{Baratella:2020lzz}
\bibitem{Baratella:2020lzz}
P.~Baratella, C.~Fernandez and A.~Pomarol,
%``Renormalization of Higher-Dimensional Operators from On-shell Amplitudes,''
Nucl. Phys. B \textbf{959} (2020), 115155.
%doi:10.1016/j.nuclphysb.2020.115155
%[arXiv:2005.07129 [hep-ph]]
%41 citations counted in INSPIRE as of 28 Nov 2023
%
%\cite{Jiang:2020mhe}
\bibitem{Jiang:2020mhe}
M.~Jiang, T.~Ma and J.~Shu,
%``Renormalization Group Evolution from On-shell SMEFT,''
JHEP \textbf{01} (2021), 101.
%doi:10.1007/JHEP01(2021)101
%[arXiv:2005.10261 [hep-ph]]
%29 citations counted in INSPIRE as of 28 Nov 2023
%
%\cite{Bern:2020ikv}
\bibitem{Bern:2020ikv}
Z.~Bern, J.~Parra-Martinez and E.~Sawyer,
%``Structure of two-loop SMEFT anomalous dimensions via on-shell methods,''
JHEP \textbf{10} (2020), 211.
%doi:10.1007/JHEP10(2020)211
%[arXiv:2005.12917 [hep-ph]]
%43 citations counted in INSPIRE as of 26 Nov 2023
%
%\cite{Baratella:2020dvw}
\bibitem{Baratella:2020dvw}
P.~Baratella, C.~Fernandez, B.~von Harling and A.~Pomarol,
%``Anomalous Dimensions of Effective Theories from Partial Waves,''
JHEP \textbf{03} (2021), 287.
%doi:10.1007/JHEP03(2021)287
%[arXiv:2010.13809 [hep-ph]]
%23 citations counted in INSPIRE as of 28 Nov 2023
%
%\cite{AccettulliHuber:2021uoa}
\bibitem{AccettulliHuber:2021uoa}
M.~Accettulli Huber and S.~De Angelis,
%``Standard Model EFTs via on-shell methods,''
JHEP \textbf{11} (2021), 221;
%doi:10.1007/JHEP11(2021)221
%[arXiv:2108.03669 [hep-th]]
%36 citations counted in INSPIRE as of 28 Nov 2023
%
%\cite{DeAngelis:2023bmd}
%\bibitem{DeAngelis:2023bmd}
S.~De Angelis and G.~Durieux,
%``EFT matching from analyticity and unitarity,''
[arXiv:2308.00035 [hep-ph]].
%4 citations counted in INSPIRE as of 08 Dec 2023
%
%\cite{EliasMiro:2021jgu}
\bibitem{EliasMiro:2021jgu}
J.~Elias Miro, C.~Fernandez, M.~A.~Gumus and A.~Pomarol,
%``Gearing up for the next generation of LFV experiments, via on-shell methods,''
JHEP \textbf{06} (2022), 126.
%doi:10.1007/JHEP06(2022)126
%[arXiv:2112.12131 [hep-ph]]
%5 citations counted in INSPIRE as of 28 Nov 2023
%
%\cite{Baratella:2022nog}
\bibitem{Baratella:2022nog}
P.~Baratella, S.~Maggio, M.~Stadlbauer and T.~Theil,
%``Two-loop infrared renormalization with on-shell methods,''
Eur. Phys. J. C \textbf{83} (2023) no.8, 751.
%doi:10.1140/epjc/s10052-023-11929-6
%[arXiv:2207.08831 [hep-th]]
%4 citations counted in INSPIRE as of 28 Nov 2023
%
%\cite{Machado:2022ozb}
\bibitem{Machado:2022ozb}
C.~S.~Machado, S.~Renner and D.~Sutherland,
%``Building blocks of the flavourful SMEFT RG,''
JHEP \textbf{03} (2023), 226.
%doi:10.1007/JHEP03(2023)226
%[arXiv:2210.09316 [hep-ph]]
%9 citations counted in INSPIRE as of 28 Nov 2023
%
%\cite{Elias-Miro:2014eia}
\bibitem{Elias-Miro:2014eia}
J.~Elias-Miro, J.~R.~Espinosa and A.~Pomarol,
%``One-loop non-renormalization results in EFTs,''
Phys. Lett. B \textbf{747} (2015), 272-280.
%doi:10.1016/j.physletb.2015.05.056
%[arXiv:1412.7151 [hep-ph]]
%55 citations counted in INSPIRE as of 28 Nov 2023
%
%\cite{Cheung:2015aba}
\bibitem{Cheung:2015aba}
C.~Cheung and C.~H.~Shen,
%``Nonrenormalization Theorems without Supersymmetry,''
Phys. Rev. Lett. \textbf{115} (2015) no.7, 071601.
%doi:10.1103/PhysRevLett.115.071601
%[arXiv:1505.01844 [hep-ph]]
%115 citations counted in INSPIRE as of 28 Nov 2023
%
%\cite{Bern:2019wie}
\bibitem{Bern:2019wie}
Z.~Bern, J.~Parra-Martinez and E.~Sawyer,
%``Nonrenormalization and Operator Mixing via On-Shell Methods,''
Phys. Rev. Lett. \textbf{124} (2020) no.5, 051601.
%doi:10.1103/PhysRevLett.124.051601
%[arXiv:1910.05831 [hep-ph]]
%38 citations counted in INSPIRE as of 28 Nov 2023
%
%\cite{Jiang:2020rwz}
\bibitem{Jiang:2020rwz}
M.~Jiang, J.~Shu, M.~L.~Xiao and Y.~H.~Zheng,
%``Partial Wave Amplitude Basis and Selection Rules in Effective Field Theories,''
Phys. Rev. Lett. \textbf{126} (2021) no.1, 011601.
%doi:10.1103/PhysRevLett.126.011601
%[arXiv:2001.04481 [hep-ph]]
%38 citations counted in INSPIRE as of 28 Nov 2023
%
%\cite{Bern:1994zx}
\bibitem{Bern:1994zx}
Z.~Bern, L.~J.~Dixon, D.~C.~Dunbar and D.~A.~Kosower,
%``one-loop n point gauge theory amplitudes, unitarity and collinear limits,''
Nucl. Phys. B \textbf{425} (1994), 217-260.
%doi:10.1016/0550-3213(94)90179-1
%[arXiv:hep-ph/9403226 [hep-ph]]
%1490 citations counted in INSPIRE as of 28 Nov 2023
%
%\cite{Bern:1994cg}
\bibitem{Bern:1994cg}
Z.~Bern, L.~J.~Dixon, D.~C.~Dunbar and D.~A.~Kosower,
%``Fusing gauge theory tree amplitudes into loop amplitudes,''
Nucl. Phys. B \textbf{435} (1995), 59-101.
%doi:10.1016/0550-3213(94)00488-Z
%[arXiv:hep-ph/9409265 [hep-ph]]
%1127 citations counted in INSPIRE as of 28 Nov 2023
%
%\cite{Britto:2004nc}
\bibitem{Britto:2004nc}
R.~Britto, F.~Cachazo and B.~Feng,
%``Generalized unitarity and one-loop amplitudes in N=4 super-Yang-Mills,''
Nucl. Phys. B \textbf{725} (2005), 275-305.
%doi:10.1016/j.nuclphysb.2005.07.014
%[arXiv:hep-th/0412103 [hep-th]]
%
%\cite{Britto:2005ha}
\bibitem{Britto:2005ha}
R.~Britto, E.~Buchbinder, F.~Cachazo and B.~Feng,
%``One-loop amplitudes of gluons in SQCD,''
Phys. Rev. D \textbf{72} (2005), 065012.
%doi:10.1103/PhysRevD.72.065012
%[arXiv:hep-ph/0503132 [hep-ph]]
%
%\cite{Britto:2006sj}
\bibitem{Britto:2006sj}
R.~Britto, B.~Feng and P.~Mastrolia,
%``The Cut-constructible part of QCD amplitudes,''
Phys. Rev. D \textbf{73} (2006), 105004.
%doi:10.1103/PhysRevD.73.105004
%[arXiv:hep-ph/0602178 [hep-ph]]
%
%\cite{Mastrolia:2009dr}
\bibitem{Mastrolia:2009dr}
P.~Mastrolia,
%``Double-Cut of Scattering Amplitudes and Stokes' Theorem,''
Phys. Lett. B \textbf{678} (2009), 246-249.
%doi:10.1016/j.physletb.2009.06.033
%[arXiv:0905.2909 [hep-ph]]
%
%\cite{Ellis:2011cr}
\bibitem{Ellis:2011cr}
R.~K.~Ellis, Z.~Kunszt, K.~Melnikov and G.~Zanderighi,
%``One-loop calculations in quantum field theory: from Feynman diagrams to unitarity cuts,''
Phys. Rept. \textbf{518} (2012), 141-250.
%doi:10.1016/j.physrep.2012.01.008
%[arXiv:1105.4319 [hep-ph]]
%\cite{Ellis:1975ap}
\bibitem{Ellis:1975ap}
J.~R.~Ellis, M.~K.~Gaillard and D.~V.~Nanopoulos,
%``A Phenomenological Profile of the Higgs Boson,''
Nucl. Phys. B \textbf{106} (1976), 292.
%doi:10.1016/0550-3213(76)90382-5
%1428 citations counted in INSPIRE as of 06 Dec 2023
%
%\cite{Shifman:1979eb}
\bibitem{Shifman:1979eb}
M.~A.~Shifman, A.~I.~Vainshtein, M.~B.~Voloshin and V.~I.~Zakharov,
%``Low-Energy Theorems for Higgs Boson Couplings to Photons,''
Sov. J. Nucl. Phys. \textbf{30} (1979), 711-716
ITEP-42-1979.
%872 citations counted in INSPIRE as of 06 Dec 2023
%

%\cite{Adler:1964um}
\bibitem{Adler:1964um}
S.~L.~Adler,
%``Consistency conditions on the strong interactions implied by a partially conserved axial vector current,''
Phys. Rev. \textbf{137} (1965), B1022-B1033.
%doi:10.1103/PhysRev.137.B1022
%578 citations counted in INSPIRE as of 01 Aug 2024

%\cite{Cheung:2021yog}
\bibitem{Cheung:2021yog}
C.~Cheung, A.~Helset and J.~Parra-Martinez,
%``Geometric soft theorems,''
JHEP \textbf{04} (2022), 011.
%doi:10.1007/JHEP04(2022)011
%[arXiv:2111.03045 [hep-th]].
%42 citations counted in INSPIRE as of 01 Aug 2024

%\cite{Balkin:2021dko}
\bibitem{Balkin:2021dko}
R.~Balkin, G.~Durieux, T.~Kitahara, Y.~Shadmi and Y.~Weiss,
%``On-shell Higgsing for EFTs,''
JHEP \textbf{03} (2022), 129.
%doi:10.1007/JHEP03(2022)129
%[arXiv:2112.09688 [hep-ph]].
%22 citations counted in INSPIRE as of 01 Aug 2024

%\cite{Bertuzzo:2023slg}
\bibitem{Bertuzzo:2023slg}
E.~Bertuzzo, C.~Grojean and G.~M.~Salla,
%``ALPs, the on-shell way,''
JHEP \textbf{05} (2024), 175.
%doi:10.1007/JHEP05(2024)175
%[arXiv:2311.16253 [hep-ph]].
%3 citations counted in INSPIRE as of 01 Aug 2024

%\cite{Georgi:1986df}
\bibitem{Georgi:1986df}
H.~Georgi, D.~B.~Kaplan and L.~Randall,
%``Manifesting the Invisible Axion at Low-energies,''
Phys. Lett. B \textbf{169} (1986), 73-78.
%doi:10.1016/0370-2693(86)90688-X
%313 citations counted in INSPIRE as of 06 Dec 2023
%
%\cite{Marciano:2016yhf}
\bibitem{Marciano:2016yhf}
W.~J.~Marciano, A.~Masiero, P.~Paradisi and M.~Passera,
%``Contributions of axionlike particles to lepton dipole moments,''
Phys. Rev. D \textbf{94} (2016) no.11, 115033.
%doi:10.1103/PhysRevD.94.115033
%[arXiv:1607.01022 [hep-ph]]
%148 citations counted in INSPIRE as of 06 Dec 2023
%
%\cite{DiLuzio:2020oah}
\bibitem{DiLuzio:2020oah}
L.~Di Luzio, R.~Gr\"ober and P.~Paradisi,
%``Hunting for $CP$-violating axionlike particle interactions,''
Phys. Rev. D \textbf{104} (2021) no.9, 095027.
%doi:10.1103/PhysRevD.104.095027
%[arXiv:2010.13760 [hep-ph]]
%24 citations counted in INSPIRE as of 30 Nov 2023
%
%\cite{Bresciani:longpaper}
\bibitem{Bresciani:longpaper}
L.~C.~Bresciani, G.~Brunello, G.~Levati, M.~Mandal, P.~Mastrolia, and P.~Paradisi,
%`` ...,''
to appear.
\end{thebibliography}
\end{document}